\documentclass[runningheads]{llncs}
\usepackage[T1]{fontenc}
\usepackage{graphicx}
\usepackage{multirow}

\begin{document}
\title{Effective-LDAM: An Effective Loss Function To Mitigate Data Imbalance for Robust Chest X-Ray Disease Classification}
%
%
\author{Sree Rama Vamsidhar S\inst{1} \and
Bhargava Satya\inst{2} \and
Rama Krishna Gorthi\inst{1}}
\authorrunning{SRV et al.}
%
\institute{Indian Institute of Technology Tirupati, Andhra Pradesh, India \and
Andhra University, Andhra Pradesh, India}
\maketitle             
%
\begin{abstract}
Deep Learning (DL) approaches have gained prominence in medical imaging for disease diagnosis. Chest X-ray (CXR) classification has emerged as an effective method for detecting various diseases. Among these methodologies, Chest X-ray (CXR) classification has proven to be an effective approach for detecting and analyzing various diseases. However, the reliable performance of DL classification algorithms is dependent upon access to large and balanced datasets, which pose challenges in medical imaging due to the impracticality of acquiring sufficient data for every disease category. To tackle this problem, we propose an algorithmic-centric approach called Effective-Label Distribution Aware Margin (E-LDAM), which modifies the margin of the widely adopted Label Distribution Aware Margin (LDAM) loss function using an effective number of samples in each class. Experimental evaluations on the COVIDx CXR dataset focus on Normal, Pneumonia, and COVID-19 classification. The experimental results demonstrate the effectiveness of the proposed E-LDAM approach, achieving a remarkable recall score of 97.81\% for the minority class (COVID-19) in CXR image prediction. Furthermore, the overall accuracy of the three-class classification task attains an impressive level of 95.26\%. 

\keywords{Chest X-Ray  \and Classification \and Imbalance \and Loss function.}
\end{abstract}
\section{Introduction}
\label{sec:intro}
Data imbalance is a significant challenge encountered in the field of medical imaging, which can impact the accuracy and effectiveness of classification algorithms. In medical imaging, the imbalance arises when certain disease categories are underrepresented compared to others in the available dataset. This imbalance hampers the development of robust and reliable models for disease detection and diagnosis.
\par With the advent of deep learning techniques there have been notable advancements in automating the analysis of medical images. However, the performance and generalizability of these models heavily depend on the availability of balanced and diverse data across different disease categories. Unfortunately, achieving a balanced distribution of medical images for each disease category is challenging due to factors like the rarity of certain diseases and limitations in data collection. Classifiers trained on imbalanced datasets tend to be biased towards the more prevalent disease categories, which can lead to inaccuracies in identifying and classifying less common diseases. In recent times, an imbalance in COVID-19 chest X-ray classification poses a significant challenge in medical imaging, impacting the accuracy and reliability of diagnostic algorithms. Chest X-ray (CXR) classification has emerged as a valuable tool for the detection and analysis of COVID-19 cases. However, due to various factors such as data availability and disease prevalence, imbalanced datasets are common in COVID-19 classification tasks.\\ 
To address the challenge of data imbalance in COVID-19 chest X-ray classification, researchers have explored various techniques and strategies. These include data over sampling, under sampling, methods \cite{ozturk2020automated,bassi2022deep,ismael2021deep,islam2020combined}, re-weighting loss functions \cite{cao2019learning,buda2018systematic}, and employing advanced algorithms like \cite{li2020robust}, to handle imbalanced distributions. The aim is to ensure equitable consideration for both the majority and minority classes, thus improving the accuracy and reliability of COVID-19 classification models.
\par While data augmentation can be beneficial in certain computer vision applications, its application to medical imaging must be approached with caution due to the unique characteristics of medical images, ethical considerations, annotation requirements, limited data availability, and regulatory concerns. Hence, in this work, we focus on an algorithmic approach to mitigate the impact of data imbalance in COVID-19 chest X-ray classification, a three-class task using the COVIDx dataset \cite{wang2020covid}.
\par In this paper, we propose an algorithmic approach to enhance the performance of diagnostic algorithms. Here, we propose a modified LDAM loss function with an effective number of class samples \cite{cui2019class} to design the margin for each class, unlike the foundation work LDAM which considers the given class sample frequencies in the dataset and is termed as Effective LDAM (E-LDAM).
\par The major limitations of the LDAM loss function are it heavily relies on the presence of well-separated class boundaries, which may be challenging to achieve in datasets with overlapping or closely related classes, potentially leading to sub-optimal performance. Also, the LDAM loss struggles to handle extremely imbalanced datasets where the minority class has significantly fewer samples, leading to challenges in accurately representing and classifying the minority class.\\
Under the re-weighting techniques, the effective number of samples in Class Balanced loss \cite{cui2019class} is a modified representation of class frequencies to provide a more accurate weighting scheme that accounts for the class imbalance. It is worth noting that the effective number of samples indirectly helps the learning objective in handling overlapping samples at the decision boundary. Therefore, in this work, we propose to modify LDAM using the effective number of samples. E-LDAM provides a robust and effective margin for the generalization of minority samples. By leveraging the classification ability of Heat Guided Convolution Neural Network (HG-CNN) \cite{sivapuram2023visal} model to predict based on the fusion of global and local features of an image and the proposed E-LDAM loss function, a reliable classification model for COVID-19 detection from CXR images was developed.  The main objective of this work is to achieve a superior recall score for COVID-19 (minority class) through enhanced loss function while also maintaining a consistent accuracy of the classification model. \\
The paper is organized as follows: Section 2 provides a review of related work in imbalanced data classification, particularly in the context of COVID-19 chest X-ray classification. In Section 3, we detail our proposed methodologies, covering data preprocessing, novel loss functions, and tailored model architectures for managing data imbalance in COVID-19 classification. Section 4 outlines our experimental setup, showcases results, and offers a thorough analysis. Finally, Section 5 presents the conclusion.

\section{Related work}
This section provides an overview of the COVID-19 CXR imbalanced data classification works.
\par The wide variety of imbalanced data classification approaches in computer vision are discussed below.\\
To achieve an unbiased classification model, the approaches are broadly categorized into two categories. They are: 1. Data-centric approaches 2. Algorithmic-centric approaches. 
\par Data-centric approaches are commonly employed to address the issue of imbalanced datasets, aiming to mitigate the disparity in the distribution of majority and minority classes. This mitigation involves two primary techniques: oversampling the minority class and undersampling the majority class. However, it's worth noting that while undersampling can help rectify bias, it also comes at the cost of discarding valuable and diverse information, ultimately resulting in a limited number of samples for model training. This limitation can be particularly problematic when dealing with deep learning models \cite{us1}.\\
On the other hand, oversampling techniques, whether through data augmentation or data synthesis, predominantly rely on the existing samples from the minority classes \cite{cui2018large,perez2017effectiveness,mikolajczyk2018data}. An attempt has been made by Kim et al. \cite{kim2020m2m} to bridge the diversity gap between majority and minority class data. However, it's essential to recognize that augmentation alone does not equate to an increase in genuinely diverse data. Instead, it may lead to overfitting on the limited number of minority samples, resulting in poor generalization, especially in scenarios of extreme class imbalance. In the context of COVID-19 detection, several studies have applied data re-sampling techniques, including works by Ozturk et al. \cite{ozturk2020automated}, Bassi et al. \cite{bassi2022deep}, and Ismael et al. \cite{ismael2021deep}.
\par Algorithmic approaches aim to mitigate the inherent bias introduced by class-wise sample size disparities within a dataset. This bias is addressed through the re-weighting of the loss function, with weights being inversely proportional to the frequency of samples in a class-specific manner, as articulated in prior research \cite{buda2018systematic}. While Categorical Cross-Entropy (CE) serves as the standard loss function for balanced datasets in classification tasks, its application to imbalanced datasets can lead to detrimental effects, particularly with respect to the minority class, potentially resulting in overfitting. To tackle dataset imbalance effectively, a body of work has emerged, either in conjunction with cross-entropy loss or independently, proposing specialized loss functions that enhance class separability. This is achieved through modifications to the similarity assessment term (final logits) or the inclusion of regularizers, yielding improved performance \cite{kornblith2020s}.\\
Some variants, such as the Class-Balanced Loss \cite{cui2019class}, introduce the concept of an "effective number" of samples to serve as alternative weights in the re-weighting process. The Focal loss \cite{lin2017focal}, on the other hand, is designed to address class imbalance by down-weighting easily classified examples. Another line of research focuses on enhancing a model's ability to distinguish between classes by incorporating margins. The margin for class $i$ is defined as the minimum distance of data belonging to the $i^{th}$ class from the decision boundary. Studies in the context of imbalanced data applications explore asymmetrical margins \cite{khan2019striking,li2019overfitting}. The Label Distribution Aware Margin loss (LDAM) \cite{cao2019learning} extends the existing soft margin loss \cite{wang2018additive} by assigning larger margins to minority classes. This learning objective promotes a simpler model for the minority class, facilitating increased generalization potential, while employing a more complex model for majority classes. Class label-dependent margins are incorporated into the CE loss function under this framework.\\
In the specific domain of COVID-19 detection, a limited number of algorithmic approaches have been explored. Wong et al. \cite{wang2020covid} introduced COVID-Net, while Li et al. \cite{li2020robust} pursued discriminative cost-sensitive learning (DSCL) approach. DSCL incorporates a conditional center loss that enables the acquisition of deep discriminative representations from chest X-ray (CXR) data, allowing for adaptive adjustment of the cost associated with misclassifying COVID-19 instances within the classes. Additionally, Al-Rakhami \cite{al2021diagnosis} proposed a hybrid architecture that combines Convolutional Neural Networks (CNNs) and Recurrent Neural Networks (RNNs) with transfer learning techniques, leveraging CNNs for feature extraction from CXR images and RNNs for classification. Transfer learning, in which knowledge transfer occurs from one task to another with minimal training, is employed in this context. Furthermore, MM Rahman et al. \cite{rahman2021hog+} introduced a deep learning model based on the histogram of oriented gradients applied to input CXR images.
\section{Proposed method}
\label{sec:proposed}
In this section, we discuss the proposed E-LDAM in great detail.
\subsection{Problem formulation}
In classification tasks, Cross entropy loss is a commonly used loss function that measures the dissimilarity between predicted class probabilities and the true class labels. It quantifies the error between the predicted probability distribution and the actual distribution using the logarithmic loss. In a k-class classification task, the CE loss function for an input sample is given as shown in Eq.\ref{eq1}.
\begin{equation}
\label{eq1}
\mathcal{L}_{ce}=-log\frac{e^{z_{y}}}{{\sum_{j=1}^{k}}e^{z_j}}
\end{equation}
Where $z_j$ denotes the $j^{th}$ index value in the predicted output vector of the classification model.
\par   However, the CE loss function alone cannot handle the imbalance in data. The works \cite{wang2018additive,wang2018cosface} proved that modifying the vanilla softmax cross-entropy loss, either by operating on CE loss term or by adding regularizers, can lead to improved performance since the class separation ability at the penultimate layer improves with the modified objectives \cite{kornblith2020s}. 
\\ \textbf{Label Distribution Aware Margin Loss (LDAM)} \cite{cao2019learning} is such a loss function that addresses the imbalance data classification. It introduces a margin-based formulation that enhances the discrimination between classes, particularly for minority classes with fewer samples. By assigning larger margins to minority classes, LDAM loss aims to alleviate the impact of data imbalance and improve their representation in the learned feature space.\\
Earlier works like Large-Margin Softmax  \cite{liu2016large}, Angular Softmax \cite{liu2017sphereface}, and Additive Margin Softmax \cite{deng2019arcface} have been proposed to minimize intra-class variation in predictions and enlarge the inter-class margin. But, in contrast to the above loss functions with class-independent margins, LDAM supports using label-dependent margins which encourage bigger margins to minority classes over majority classes. The loss function equation of LDAM is given below.
\begin{equation}
\label{eq3}
\mathcal{L}_{ldam}=-log\frac{e^{z_{y}-\Delta_{y}}}{e^{z_{y}-\Delta_{y}}+\sum_{j\neq y}e^{z_j}}
\end{equation}
where, for some constant C and number of samples present in class $j$ i.e. $n_j$,
\begin{equation}
 \Delta_j=\frac{C}{n_j^{1/4}} \hspace{0.2cm} for \hspace{0.2cm} j \in \{1,...k\} 
\end{equation}
\par Further, under reweighting techniques, \textbf{Class Balanced loss (CB loss)} \cite{cui2019class} was proposed to mitigate the bias towards majority classes in imbalanced datasets. It dynamically adjusts the loss contribution of each class based on their respective frequencies to provide equal consideration to all classes during training. By reweighting the loss function, class-balanced loss ensures that each class contributes proportionally to the overall loss, effectively addressing the imbalance issue. CB loss is a re-weighting approach, specifically proposed to avoid the problem of data overlapping within the class. The effective number of samples can be described as a volume of a set of unique samples. The effective number of samples for each class could be calculated from the following equation.
\begin{equation}
E_n=\frac{1-{\beta^n}}{1-{\beta}}
\end{equation} 
where, \textit{n} is the sample size of a particular class, and $\beta \in[0, 1)$ is a hyper-parameter. \\
While cross-entropy loss is a standard choice, LDAM loss and class-balanced loss offer specific adaptations to handle imbalanced datasets. The incorporation of these loss functions in classification tasks helps to achieve fair and accurate predictions, particularly in scenarios where class distribution is uneven. We note a significant distinction between the re-weighting approach and LDAM. The scalar factor introduced in re-weighting is solely dependent on the class, while in LDAM, it also depends on the model's output. 
\subsection{Effective-LDAM (E-LDAM)}
 To optimize the decision boundary based on the effective number of samples in a class and mitigate the influence of similar samples, we propose a modified LDAM loss function. Instead of using the actual number of samples, we calculate the loss by considering the effective number of samples. Unlike the existing works \cite{cui2019class,cao2019learning} etc, which use $E_n$ to re-weight the learning objective, the proposed approach makes use of $E_n$ to design the decision boundary. By incorporating the concept of the effective number of samples, the LDAM loss function exhibits improved recall scores and precision for minority classes. This adjustment favors the minority classes by shifting the decision boundary optimally in favor of the minority class. The enhanced LDAM loss function, incorporating the effective number of samples, can be formulated as shown in the below equation.
 \begin{equation}
\mathcal{L}_{e-ldam}=-log\frac{e^{z_{y}-\Delta_{y}}}{e^{z_{y}-\Delta_{y}}+\sum_{j\neq y}e^{z_j}}
\end{equation}
\begin{equation}
 \Delta_j=\frac{C}{{E_{n_j}}^{1/r}} \hspace{0.2cm} for \hspace{0.2cm} j \in \{1,...k\} 
\end{equation}
$E_n$ is the effective number of samples of class \textit{n}, where $E_n$ could be calculated by using  $\frac{1-{\beta^n}}{1-{\beta}}$ and $r$ is a positive integer.

\section{Experimental results}
\label{exp}
\subsection{Architecture details} 
\begin{figure*}[h]
        \centering
        \includegraphics[width=\textwidth, height = 6cm]{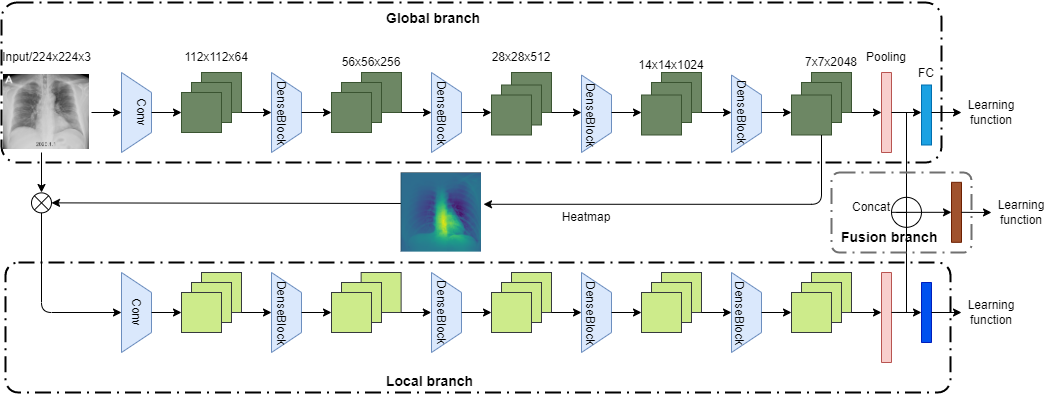}
        \caption{\centering  HG-CNN for Chest X-ray image classification.}
        \label{fig:arch}
    \end{figure*}

In the experiments,  we utilized a convolutional neural network model guided by heatmaps, consisting of three branches: global, local, and fusion. This model is known as the Heat guided Convolutional Neural Network (HG-CNN). It is designed to detect infections, especially those that appear as localized issues like COVID-19.
 HG-CNN combines information from the whole image and specific areas to improve accuracy and reduce errors caused by noise. HG-CNN uses a well-known architecture called DenseNet121 and operates in three stages, with each stage classifying images into one of three categories: COVID-19, Pneumonia, or Normal. The HG-CNN model's structure is given in Figure \ref{fig:arch}.

\par In the first stage (Stage 1), we prepare the input chest X-ray images by making them a standard size of 224x224 pixels and improving their quality using the histogram equalization technique. Then, we feed the processed image into the global branch for the three class classification i.e. Normal, Pneumonia, and COVID-19. From the trained global branch, we extract a heap map showing where the model should pay attention and the feature vector from the penultimate layer of dimension $1024 \times 1$. We resize and multiply the heap map with the original image and pass it to the local branch for the next stage (Stage 2). In Stage 2, the local branch produces another 1024-value feature vector at the penultimate layer. During training at the local branch, we focus on the parts of the image that matter the most, thanks to the attention map.\\
In the final stage (Stage 3), we combine the two feature vectors from Stages 1 and 2. We do this by putting them through a single layer called the fusion branch. This layer then tells us the probabilities of the image belonging to COVID-19, Pneumonia, or being Normal. It's essential to note that we train this model in stages: first the global branch, then the local branch, and finally, the fusion branch.\\
\textbf{Working mechanism}:
The way HG-CNN works is a lot like how a radiologist reads X-rays. First, we train the global branch, which is similar to a radiologist taking an initial look at the entire X-ray image. Then, we use heatmaps to find specific regions of interest, like areas with lesions, and include them in the X-ray image, making it weighted. This weighted image goes into the local branch, which is similar to a radiologist zooming in on the problem area after the initial assessment. Finally, we put together the global and local information to make a final decision, just like a radiologist who considers both the big picture and the details before making a diagnosis.\\
Also, it's important to mention that we train the model in two steps. First, we train the HG-CNN on the National Institutes of Health (NIH) dataset. Then, we fine-tune it for the specific 3-class classification task using the COVIDx dataset. We carefully adjust key settings like the number of training cycles, learning rates, how fast the learning rate decreases, and how many examples the model looks at in each training round, based on experiments at each stage of development.
\subsection {Dataset}
\label{sec:Details about dataset}
The dataset utilized for training and evaluating the proposed model comprises a total of 19,364 Chest X-ray (CXR) images sourced from the COVIDx dataset (version 8) by Wang et al. \cite{wang2020covid}. The COVIDx dataset is a compilation originating from five publicly available data repositories: (1) COVID-19 Image Data Collection, (2) ActualMed COVID-19 Chest X-ray Dataset Initiative \cite{wang2020actualmed}, (3) RSNA Pneumonia Detection Challenge dataset, which utilized publicly available CXR data, (4) COVID-19 radiography database, and (5) RSNA International COVID-19 Open Radiology Database (RICORD) as reported in \cite{tsai2021rsna}. It's important to note that the dataset exhibits an imbalanced distribution of CXR images across various classes, as inherited from the COVIDx dataset. The distribution of samples for each class is detailed in Table \ref{tab:dataset}.
\begin{table}[!h]
\begin{center}
\caption{Sample distribution among the classes in COVIDx dataset.}
\label{tab:dataset}
\begin{tabular}{ |p{2cm}||p{1.8cm}|p{1.8cm}|p{1.8cm}|  }
\hline
 \multicolumn{4}{|c|}{COVIDx Dataset} \\
 \hline
 \hline
  & \textbf{COVID-19} & \textbf{Pneumonia} & \textbf{Normal}\\
 \hline
 Train split & 4649 & 5964 & 8751 \\ \cline{1-4}
 Test spilt &  274 & 105  & 100\\ \cline{1-4}
 Total  & \textbf{4923} &  \textbf{6069} & \textbf{8581} \\ \cline{1-4}
 \hline
 \end{tabular}
 \end{center}

 \end{table}

\subsection {Results}
\label{sec:Details about results} 
This section consists of details and results of experiments performed with the HG-CNN model on different loss functions. This section also includes the comparative results of various works with the proposed approach. The below Table 2. consists of the experiment results of each stage in the HG-CNN model Global, Local, and Fusion branches. Every stage in the HG-CNN model is experimented with different loss functions. The table below contains the overall accuracy of each stage and the recall score of COVID-19 obtained from experiments. From Table \ref{tab:HG-CNN}, we can also observe that effective loss functions have exhibited a better performance when compared to standard loss functions. The effective loss functions have increased the model's capability in handling the minority class COVID-19 at each phase of the model. They are able to improve the recall score for COVID-19 while maintaining the overall accuracy of the model.

\begin{table}[!h]

\begin{center}
\caption {Results from the experiments performed on COVIDx dataset with proposed HG-CNN model across the standard loss functions \& the proposed loss function.}
\label{tab:HG-CNN}
\begin{tabular}{|p{2cm}|p{2cm}|p{1.5cm}|p{1.5cm}|p{1.5cm}|  }
 \hline
   \textbf{Loss} \hspace{1cm} \textbf{function} & \textbf{Metric} & \textbf{Global Branch} & \textbf{Local Branch} & \textbf{Fusion Branch}\\
 \hline
  \hline
 \multirow{2}{*}{CE} & Accuracy & 91.86\% & 87.27\%  & 93.74\%\\ \cline{2-5}
                    & Recall &  90.51\% & 88.69\%  & 93.43\%\\ \cline{2-5}
 \hline
  \hline
 \multirow{2}{*}{CB-CE} & Accuracy & 92.28\% & 84.76\%  & 92.28\%\\ \cline{2-5}
   & Recall &  93.43\% & 88.32\%  & 95.26\%\\ \cline{2-5}
 \hline
  \hline
 \multirow{2}{*}{LDAM} &Accuracy & 93.32\% & 86.22\%  & 92.69\%\\ \cline{2-5}
  & Recall &  94.16\% & 86.86\%  & 95.26\%\\ \cline{2-5}
 \hline
  \hline
 \multirow{2}{*}{E-LDAM} &Accuracy & 95\% & 88.10\%  & 95.82\%\\ \cline{2-5}
  & Recall &  97.08\% & 87.96\%  & 97.81\% \\ \cline{1-4}
 \hline

\end{tabular}
\end{center}

\end{table}

E-LDAM loss function with the HG-CNN model has outperformed the various other significant contemporary works in detecting COVID-19 infection. The  E-LDAM loss has enhanced the model's capability in detecting in handling the minority classes and has increased the recall score for the minority class like COVID-19 while maintaining a good accuracy when compared to other approaches. The comparative results of the proposed approach with various other approaches are shown in Table 3. The above table consists of other works in detecting COVID-19 infection by handling the imbalance in the dataset with algorithmic approaches. 
\begin{table}[!h]
\begin{center}
\caption {Comparison results on COVID-19 detection.}
\begin{tabular}{|p{2.75cm}|p{3.2cm}|p{1.8cm}|p{1.8cm}|}\hline
\textbf{Method} & \textbf{Technique} & \textbf{Accuracy} & \textbf{Recall}   \\
\hline
Li, et al. \cite{li2020robust}  & Data augmentation &97.01\% &97\%\\
\hline
Rahman et al. \cite{rahman2021hog+} & Data augmentation & 96.74\% &96.5\%\\
\hline
Proposed method & Loss function & 95.82\% &97.81\%\\
\hline
\end{tabular}
\end{center}

\end{table}

\section{Conclusion}
\label{sec:Conclusion}

In conclusion, our work introduces a novel loss function, E-LDAM, which incorporates the effective number of samples as weights for classification margin design, thereby enhancing the performance of a CXR classification model based on Heat Guided Convolutional Neural Network. Our study demonstrates that E-LDAM outperforms the existing LDAM loss function by $>3\%$ in recall specifically for COVID-19 CXR images and $>2\%$ in overall accuracy. These findings underscore the effectiveness of E-LDAM in mitigating data imbalance challenges, improving disease diagnosis accuracy in medical imaging, and facilitating the development of more reliable diagnostic systems.  We aim to further advance our work by exploring additional algorithmic approaches and robust architectures to enhance the diagnosis of various data-scarce infections in the medical domain. 

%
%
\bibliographystyle{splncs04}
\bibliography{refer.bib}

\begin{thebibliography}{10}
\providecommand{\url}[1]{\texttt{#1}}
\providecommand{\urlprefix}{URL }
\providecommand{\doi}[1]{https://doi.org/#1}

\bibitem{al2021diagnosis}
Al-Rakhami, M.S., Islam, M.M., Islam, M.Z., Asraf, A., Sodhro, A.H., Ding, W.: Diagnosis of covid-19 from x-rays using combined cnn-rnn architecture with transfer learning. MedRxiv pp. 2020--08 (2021)

\bibitem{bassi2022deep}
Bassi, P.R., Attux, R.: A deep convolutional neural network for covid-19 detection using chest x-rays. Research on Biomedical Engineering  \textbf{38}(1),  139--148 (2022)

\bibitem{buda2018systematic}
Buda, M., Maki, A., Mazurowski, M.A.: A systematic study of the class imbalance problem in convolutional neural networks. Neural networks  \textbf{106},  249--259 (2018)

\bibitem{cao2019learning}
Cao, K., Wei, C., Gaidon, A., Arechiga, N., Ma, T.: Learning imbalanced datasets with label-distribution-aware margin loss. arXiv preprint arXiv:1906.07413  (2019)

\bibitem{cui2019class}
Cui, Y., Jia, M., Lin, T.Y., Song, Y., Belongie, S.: Class-balanced loss based on effective number of samples. In: Proceedings of the IEEE/CVF conference on computer vision and pattern recognition. pp. 9268--9277 (2019)

\bibitem{cui2018large}
Cui, Y., Song, Y., Sun, C., Howard, A., Belongie, S.: Large scale fine-grained categorization and domain-specific transfer learning. In: Proceedings of the IEEE conference on computer vision and pattern recognition. pp. 4109--4118 (2018)

\bibitem{deng2019arcface}
Deng, J., Guo, J., Xue, N., Zafeiriou, S.: Arcface: Additive angular margin loss for deep face recognition. In: Proceedings of the IEEE/CVF conference on computer vision and pattern recognition. pp. 4690--4699 (2019)

\bibitem{us1}
He, H., Garcia, E.A.: Learning from imbalanced data. IEEE Transactions on Knowledge and Data Engineering  \textbf{21}(9),  1263--1284 (2009). \doi{10.1109/TKDE.2008.239}

\bibitem{islam2020combined}
Islam, M.Z., Islam, M.M., Asraf, A.: A combined deep cnn-lstm network for the detection of novel coronavirus (covid-19) using x-ray images. Informatics in medicine unlocked  \textbf{20},  100412 (2020)

\bibitem{ismael2021deep}
Ismael, A.M., {\c{S}}eng{\"u}r, A.: Deep learning approaches for covid-19 detection based on chest x-ray images. Expert Systems with Applications  \textbf{164},  114054 (2021)

\bibitem{khan2019striking}
Khan, S., Hayat, M., Zamir, S.W., Shen, J., Shao, L.: Striking the right balance with uncertainty. In: Proceedings of the IEEE/CVF Conference on Computer Vision and Pattern Recognition. pp. 103--112 (2019)

\bibitem{kim2020m2m}
Kim, J., Jeong, J., Shin, J.: M2m: Imbalanced classification via major-to-minor translation. In: Proceedings of the IEEE/CVF conference on computer vision and pattern recognition. pp. 13896--13905 (2020)

\bibitem{kornblith2020s}
Kornblith, S., Lee, H., Chen, T., Norouzi, M.: What's in a loss function for image classification? arXiv preprint arXiv:2010.16402  (2020)

\bibitem{li2020robust}
Li, T., Han, Z., Wei, B., Zheng, Y., Hong, Y., Cong, J.: Robust screening of covid-19 from chest x-ray via discriminative cost-sensitive learning. arXiv preprint arXiv:2004.12592  (2020)

\bibitem{li2019overfitting}
Li, Z., Kamnitsas, K., Glocker, B.: Overfitting of neural nets under class imbalance: Analysis and improvements for segmentation. In: International Conference on Medical Image Computing and Computer-Assisted Intervention. pp. 402--410. Springer (2019)

\bibitem{lin2017focal}
Lin, T.Y., Goyal, P., Girshick, R., He, K., Doll{\'a}r, P.: Focal loss for dense object detection. In: Proceedings of the IEEE international conference on computer vision. pp. 2980--2988 (2017)

\bibitem{liu2017sphereface}
Liu, W., Wen, Y., Yu, Z., Li, M., Raj, B., Song, L.: Sphereface: Deep hypersphere embedding for face recognition. In: Proceedings of the IEEE conference on computer vision and pattern recognition. pp. 212--220 (2017)

\bibitem{liu2016large}
Liu, W., Wen, Y., Yu, Z., Yang, M.: Large-margin softmax loss for convolutional neural networks. arXiv preprint arXiv:1612.02295  (2016)

\bibitem{mikolajczyk2018data}
Miko{\l}ajczyk, A., Grochowski, M.: Data augmentation for improving deep learning in image classification problem. In: 2018 international interdisciplinary PhD workshop (IIPhDW). pp. 117--122. IEEE (2018)

\bibitem{ozturk2020automated}
Ozturk, T., Talo, M., Yildirim, E.A., Baloglu, U.B., Yildirim, O., Acharya, U.R.: Automated detection of covid-19 cases using deep neural networks with x-ray images. Computers in biology and medicine  \textbf{121},  103792 (2020)

\bibitem{perez2017effectiveness}
Perez, L., Wang, J.: The effectiveness of data augmentation in image classification using deep learning. arXiv preprint arXiv:1712.04621  (2017)

\bibitem{rahman2021hog+}
Rahman, M.M., Nooruddin, S., Hasan, K., Dey, N.K.: Hog+ cnn net: Diagnosing covid-19 and pneumonia by deep neural network from chest x-ray images. SN computer science  \textbf{2}(5),  1--15 (2021)

\bibitem{sivapuram2023visal}
Sivapuram, A.K., Ravi, V., Senthil, G., Gorthi, R.K., et~al.: Visal—a novel learning strategy to address class imbalance. Neural Networks  \textbf{161},  178--184 (2023)

\bibitem{tsai2021rsna}
Tsai, E.B., Simpson, S., Lungren, M.P., Hershman, M., Roshkovan, L., Colak, E., Erickson, B.J., Shih, G., Stein, A., Kalpathy-Cramer, J., et~al.: The rsna international covid-19 open radiology database (ricord). Radiology  \textbf{299}(1), ~E204 (2021)

\bibitem{wang2018additive}
Wang, F., Cheng, J., Liu, W., Liu, H.: Additive margin softmax for face verification. IEEE Signal Processing Letters  \textbf{25}(7),  926--930 (2018)

\bibitem{wang2018cosface}
Wang, H., Wang, Y., Zhou, Z., Ji, X., Gong, D., Zhou, J., Li, Z., Liu, W.: Cosface: Large margin cosine loss for deep face recognition. In: Proceedings of the IEEE conference on computer vision and pattern recognition. pp. 5265--5274 (2018)

\bibitem{wang2020actualmed}
Wang, L., Wong, A., Lin, Z., McInnis, P., Chung, A., Gunraj, H., Lee, J., Ross, M., VanBerlo, B., Ebadi, A., et~al.: Actualmed covid-19 chest x-ray dataset initiative. URL: https://github. com/agchung/Actualmed-COVID-chestxraydataset  (2020)

\bibitem{wang2020covid}
Wang, L., Lin, Z.Q., Wong, A.: Covid-net: A tailored deep convolutional neural network design for detection of covid-19 cases from chest x-ray images. Scientific Reports  \textbf{10}(1),  1--12 (2020)

\end{thebibliography}

\end{document}